\documentclass[12pt,epsf]{article}
\usepackage{graphicx}
\usepackage{epsfig}
\begin{document}

\def\a{\alpha}
\def\b{\beta}
\def\c{\varepsilon}
\def\d{\delta}
\def\e{\epsilon}
\def\f{\phi}
\def\g{\gamma}
\def\h{\theta}
\def\k{\kappa}
\def\l{\lambda}
\def\m{\mu}
\def\n{\nu}
\def\p{\psi}
\def\q{\partial}
\def\r{\rho}
\def\s{\sigma}
\def\t{\tau}
\def\u{\upsilon}
\def\v{\varphi}
\def\w{\omega}
\def\x{\xi}
\def\y{\eta}
\def\z{\zeta}
\def\D{\Delta}
\def\G{\Gamma}
\def\H{\Theta}
\def\L{\Lambda}
\def\F{\Phi}
\def\P{\Psi}
\def\S{\Sigma}

\def\o{\over}
\def\beq{\begin{eqnarray}}
\def\eeq{\end{eqnarray}}
\newcommand{\gsim}{ \mathop{}_{\textstyle \sim}^{\textstyle >} }
\newcommand{\lsim}{ \mathop{}_{\textstyle \sim}^{\textstyle <} }
\newcommand{\vev}[1]{ \left\langle {#1} \right\rangle }
\newcommand{\bra}[1]{ \langle {#1} | }
\newcommand{\ket}[1]{ | {#1} \rangle }
\newcommand{\EV}{ {\rm eV} }
\newcommand{\KEV}{ {\rm keV} }
\newcommand{\MEV}{ {\rm MeV} }
\newcommand{\GEV}{ {\rm GeV} }
\newcommand{\TEV}{ {\rm TeV} }
\def\diag{\mathop{\rm diag}\nolimits}
\def\Spin{\mathop{\rm Spin}}
\def\SO{\mathop{\rm SO}}
\def\O{\mathop{\rm O}}
\def\SU{\mathop{\rm SU}}
\def\U{\mathop{\rm U}}
\def\Sp{\mathop{\rm Sp}}
\def\SL{\mathop{\rm SL}}
\def\tr{\mathop{\rm tr}}

\def\IJMP{Int.~J.~Mod.~Phys. }
\def\MPL{Mod.~Phys.~Lett. }
\def\NP{Nucl.~Phys. }
\def\PL{Phys.~Lett. }
\def\PR{Phys.~Rev. }
\def\PRL{Phys.~Rev.~Lett. }
\def\PTP{Prog.~Theor.~Phys. }
\def\ZP{Z.~Phys. }

\newcommand{\bear}{\begin{array}}  
\newcommand {\eear}{\end{array}}
\newcommand{\la}{\left\langle}  
\newcommand{\ra}{\right\rangle}
\newcommand{\non}{\nonumber}  
\newcommand{\ds}{\displaystyle}
\newcommand{\red}{\textcolor{red}}
\def\ubl{U(1)$_{\rm B-L}$}
\def\REF#1{(\ref{#1})}
\def\lrf#1#2{ \left(\frac{#1}{#2}\right)}
\def\lrfp#1#2#3{ \left(\frac{#1}{#2} \right)^{#3}}
\def\OG#1{ {\cal O}(#1){\rm\,GeV}}


\baselineskip 0.7cm

\begin{titlepage}

\begin{flushright}
UT-10-08\\
IPMU 10-0079
\end{flushright}

\vskip 1.35cm
\begin{center}
{\large \bf
LHC Reach of Low Scale Gauge Mediation with Perturbatively Stable Vacuum 
}
\vskip 1.2cm
Ryosuke Sato and Satoshi Shirai 
\vskip 0.4cm

{\it  Department of Physics, University of Tokyo,\\
   Tokyo 113-0033, Japan\\
 Institute for the Physics and Mathematics of the Universe (IPMU), 
University of Tokyo,\\ Chiba 277-8568, Japan}

\vskip 1.5cm

\abstract{
Very light gravitino scenario $m_{3/2}\lsim 16$ eV is very interesting, since there is no cosmological problem.
However in such a scenario, stability of the vacuum is an important issue.
Recently, Yonekura and one of the authors RS have investigated the parameter space of a low scale gauge mediation
with a perturbatively stable vacuum and found that there are severe upper bounds on the gaugino masses.
In this Letter, we show that such a model can be completely excluded/discovered at very early stage of the LHC run.
 }
\end{center}
\end{titlepage}

\setcounter{page}{2}

\section{Introduction}\label{sec:1}
Low scale gauge mediation is attractive,
because it can achieve a very light gravitino mass $m_{3/2} \lsim 16~\EV$,
which satisfies all constraints from cosmology~\cite{Viel:2005qj}.
However in this scenario, stability of the vacuum is an important issue~\cite{Hisano:2007gb}.
It is known that there are two kinds of model which can have the light gravitino and a sufficiently stable vacuum.
One is the model of Refs. \cite{Izawa:1997gs}
using the Izawa-Yanagida-Intriligator-Thomas (IYIT) model \cite{Izawa:1996pk} as a SUSY breaking sector,
and the other is the model of Refs. \cite{Kitano:2006xg}
gauging a flavor symmetry of the Intriligator-Seiberg-Shih (ISS) model \cite{Intriligator:2006dd}.
The SUSY breaking vacuum of the IYIT model is absolutely stable.
Although the vacuum of the ISS model is only perturbatively stable, the lifetime of the vacuum is independent of the gravitino mass.
Therefore, these models can have both the light gravitino and the sufficiently stable vacuum at the same time.

However, in these two models,
it was observed that the Minimal Supersymmetric Standard Model (MSSM) gaugino masses are suppressed.
Therefore, these models are severely constrained by the Tevatron bound~\cite{Aaltonen:2009tp,Meade:2009qv}.
Recently, Yonekura and one of the present authors RS have studied the parameter space of the low scale gauge mediation model
which has a stable SUSY breaking vacuum~\cite{Sato:2009dk}.
As the result, they have found almost all the parameter space is excluded by the Tevatron bound
when the gravitino mass is less than $16~\EV$.
However, these models have not been completely excluded.
To exclude them completely,
we discuss the discovery region at early stage of the LHC run.
As the result, 
we show that such a model can be completely excluded/discovered at very early stage of the LHC run
even if we relax the upper bound of the gravitino mass to $32~\EV$.

This Letter is organized as follows.
In section \ref{sec:model}, we review the model discussed in Ref.~\cite{Sato:2009dk}.
In section \ref{sec:signature}, we estimate the discovery region of the low scale gauge mediation at the LHC.

\section{Model}\label{sec:model}
In this section, we review two kinds of the low scale gauge mediation model.

\subsection*{Model A (based on the IYIT model)}

First, we review the model \cite{Izawa:1997gs} based on the IYIT model \cite{Izawa:1996pk} as a SUSY breaking sector.
At low energies, the effective superpotential of the SUSY breaking sector\footnote{
For concreteness, we take the IYIT model as the SUSY breaking sector,
but one can take other SUSY breaking models of which the effective superpotential is given by Eq.~(\ref{eq:eff}).
}
is given by
\begin{eqnarray}
W \simeq f Z. \label{eq:eff}
\end{eqnarray}
Here, $Z$ is a singlet chiral superfield.
The SUSY breaking vacuum of the IYIT model is stable.
We introduce $N_F$ flavors of messenger multiplets which transform
as {\bf 5} and ${\bf {\bar 5}}$ under the Grand Unified Theory (GUT) gauge group ${\rm SU}(5)_{{\rm GUT}}$.
Large $N_F$ is desirable to increase the MSSM gaugino masses.
However, the perturbative gauge coupling unification is lost when $N_F \geq 5$~\cite{Jones:2008ib}.
Therefore, we choose $N_F=4$.
We will write the messenger quark multiplets as $\Psi_{d,i},{\tilde \Psi_{d,i}}~(i=1,\cdots,4)$
and the messenger lepton multiplets as $\Psi_{l,i},{\tilde \Psi_{l,i}}~(i=1,\cdots,4)$.
We assume that the superpotential of the messenger sector and the SUSY breaking sector is given as follows :
\begin{eqnarray}
W &=& f Z + \sum_{\chi = d,l} \sum_{i,j} M_{\chi,ij}(Z) {\tilde \Psi}_{\chi,i} \Psi_{\chi,j},\\
&& (M_{\chi,ij}(Z) = m_{\chi,ij} + k_{\chi,ij} Z).
\end{eqnarray}
The stability of the SUSY breaking vacuum requires $\det (m+kZ)=\det m$ \cite{Komargodski:2009jf} and $km^{-1}k=0$ \cite{Sato:2009dk}.
According to Ref. \cite{Sato:2009dk}, the gaugino masses are maximized when $M_\chi(Z)$ is given by
\begin{eqnarray}
M_\chi(Z) = \left(
\begin{array}{cccc}
k_\chi Z&m_\chi&&  \\
m_\chi&&& \\
&&k_\chi Z&m_\chi\\
&&m_\chi&
\end{array}
\right). \label{eq:matrixelement}
\end{eqnarray}
This model has an R-symmetry. Then, we assume the R-symmetry is spontaneously broken to gain the gaugino masses.
We do not specify the mechanism generating the R-symmetry breaking VEV $\la Z\ra$, and treat $\la Z\ra$ as a free parameter.
The parameter $f$ is determined by $f=\sqrt{ \sqrt{3} m_{3/2} M_{\rm Pl} }$,
where $M_{\rm Pl} \simeq 2.4 \times 10^{18}~\GEV$ is the reduced Planck mass.
After all, we have six parameters, $k_d,~k_l,~m_d,~m_l,~\la Z\ra$ and $m_{3/2}$ in the model.

We have some comments on the possible region of these parameters.
The Yukawa interaction is not asymptotic free.
Then, if we require the Yukawa interaction of the messenger sector is perturbative up to the GUT scale $2\times 10^{16}~\GEV$,
there are upper bounds of the Yukawa coupling constants $k_d$ and $k_l$.
See Fig. 2 in Ref.~\cite{Sato:2009dk}.
The condition that the Standard Model gauge symmetry is not broken, i.e.,
the messenger scalar VEV $\la \Psi_\chi \ra = \la \tilde\Psi_\chi \ra=0$ requires that $m_\chi \geq \sqrt{k_\chi f}$.

\subsection*{Model B (based on the ISS model)}

Next, we review the model using the ISS model \cite{Intriligator:2006dd}.
The SUSY breaking vacuum of the ISS model is meta-stable.
The vacuum can be sufficiently stable if the cutoff scale $\L_{\rm cut}$ is much larger than the messenger mass scale $M_{\rm mes}$.
However, to gain the large gaugino masses comparable to ones of the model A,
this model requires very low cutoff scale $\L_{\rm cut}/M_{\rm mes} \lsim {\cal O}(10)$ \cite{Sato:2009dk}.
This means large vacuum tunneling rate.
Therefore, it is doubtful whether we can get the large gaugino masses when we require the stability of the SUSY breaking vacuum.
See Ref.~\cite{Sato:2009dk} for details.

In both of the models, The sfermions are much heavier than the gauginos.
Then, the role of the gaugino masses are important when we discuss the discovery region.
Because the gaugino masses in the model B are smaller than the model A,
the model B are excluded/discovered if the model A are excluded/discovered.
In the following of the paper, we discuss the discovery region of the model A.

\section{LHC signature}\label{sec:signature}

In this section, we estimate the discovery region at $\sqrt{s}=7$ TeV 
and integrated luminosity ${\cal L}={\cal O}(1)~{\rm fb}^{-1}$.
We have used the programs Pythia 6.4.19 \cite{Sjostrand:2006za} and fast detector simulation AcerDET-1.0 \cite{RichterWas:2002ch}.
In the fast simulation, the detection efficiency of a photon which passes a certain isolation criteria is 100 \%.
However as discussed in Ref.~\cite{Aad:2009wy}, full simulation result indicates lower efficiency.
Hereafter we assume selection efficiency of the isolated photon with $p_{\rm T}>20$ GeV is 65 \%.

In the light gravitino scenario discussed in the previous section,
one of important features of LHC signature is prompt decay of the next to lightest supersymmetric particle (NLSP) into
the gravitino.
The decay length of the NLSP is written as
\beq
c\tau_{\rm NLSP} \simeq 18~\mu{\rm m} \left(\frac{m_{3/2}}{1~\EV}\right)^2\left(\frac{m_{\rm NLSP}}{100~\GEV}\right)^{-5},
\eeq
which is much smaller than the detector size.
Therefore all MSSM particles decay promptly.

\subsection*{Neutralino NLSP}

In almost all the parameter region of the present model, the NLSP is a bino-like neutralino.
Let us comment on the other possibilities for the NLSP.
In some parameter region, the NLSP is a higgsino-like or wino-like neutralino,
but the mass of the NLSP is small, $m_{\tilde \chi_1^0} \lsim 90~\GEV$.
This region is excluded by the CDF bound~\cite{Meade:2009qv}.
Also, there is a region where the lightest chargino is the NLSP,
but the mass is small, $m_{\tilde \chi_1^\pm} \lsim 90~\GEV$.
This region is excluded by the LEP bound~\cite{Amsler:2008zzb}.
Finally, it is possible to realize the gluino NLSP.
In this case, the expected collider signature is di-jet and missing energy.
However, the upper bound on the gluino NLSP is about $200~\GEV$ and it is excluded by the Tevatron di-jet search~\cite{Aaltonen:2009xp}.
Hereafter, we assume that the NLSP is a bino-like neutralino.

\subsection*{Event selection}

The prompt decay of the lightest neutralino $\tilde{\chi}^0_1 \to \gamma \tilde{G}_{3/2}$ gives
very strong clue for the SUSY discovery, i.e., high $p_{\rm T}$ photons and missing energy.
The gluino pair production $pp\to\tilde{g}\tilde{g}$ 
and wino-like neutralino and chargino production  $pp\to \tilde{\chi}\tilde{\chi}$ are dominant SUSY production.
The produced gluino decays into the lighter SUSY particle with high energy jets.
For this mode, multi-jets and large missing energy are expected in addition to the photon signal.
Following Ref.~\cite{Aad:2009wy}, we impose the following cuts
\begin{itemize}
\item At least two isolated photons with $p_{\rm T}>20$ GeV.
\item At least four jets with $p_{\rm T}>50$ GeV.
\item The leading jet with $p_{\rm T}>100$ GeV.
\item $E_{\rm T, miss}> \max(100~{\rm GeV}, 0.2M_{\rm eff})$, where
\beq
M_{\rm eff} \equiv \sum_{\rm 4~leading~jets} p_{{\rm T},j}+E_{\rm T, miss}+\sum_{\rm leptons} p_{\rm T,\ell}.
\eeq
\end{itemize}
After these cuts, the remaining SM background is $t\bar{t}$ events.
According to Ref.~\cite{Aad:2009wy}, the cross section of the background events is 0.1 fb for $\sqrt{s}=14$ TeV.
In the case of the $\sqrt{s}=7$ TeV, the cross section of this background will be suppressed.
We assume conservatively its cross section is 0.1 fb for $\sqrt{s}=7$ TeV.

We consider another discovery mode, which requires harder photons.
The following cuts are imposed:
\begin{itemize}
\item At least two isolated photons with $p_{\rm T, \gamma1}>60$ GeV and $p_{\rm T, \gamma2}>30~\GEV$.
\item $E_{\rm T, miss}> 80$ GeV.
\item $m_{\gamma_1\gamma_2}> 100$ GeV.
\end{itemize}
After this cut, the main SM background (BG) is $t\bar{t} +n\gamma$'s, $\gamma \gamma$ and $\gamma W(\to e\nu)$ in which
an electron from $W$ boson decay is misidentified as a photon.
To estimate the cross section of the background, we use the program MadGraph 4.4 \cite{Alwall:2007st}.
In order to make correction from the fake rate of electron, NLO effects and detailed detector performance,
we multiply the BG cross section by an overall scale factor to reproduce the result of \cite{AdamBourdarios:2009zz}.
We estimate this cross section of $t\bar{t}+n\gamma$'s is 0.2 fb,
$\gamma\gamma$ 0.1 fb,
$\gamma W(\to e\nu)$ 0.5 fb.\footnote{
This estimation is rough.
However, we do not have to know the precise value,
because the BG is small for ${\cal L} = {\cal O}(1)~{\rm fb}^{-1}$.
}
\subsection*{Parameter Search}
In the gauge-mediated supersymmetry breaking (GMSB) model, the MSSM gaugino masses at the messenger scale are parametrized as:
 \beq \label{eq:gaugino-mass}
 M_a = \frac{\a_a}{4\pi} \L_{Ga}~~~(a=1,2,3),
 \eeq
and sfermion masses
\beq
m^2_{\f_i}= \left(\frac{\a_1}{4\pi}\right)^2 C_1(i)\Lambda_{S1}^2+\left(\frac{\a_2}{4\pi}\right)^2
C_2(i)\Lambda_{S2}^2+\left(\frac{\a_3}{4\pi}\right)^2C_3(i)\Lambda_{S3}^2.
\eeq
In the present model,
\beq
\Lambda_{G1}=\frac{3}{5}\Lambda_{G2}+\frac{2}{5}\Lambda_{G3},\\
\Lambda_{S1}^2=\frac{3}{5}\Lambda_{S2}^2+\frac{2}{5}\Lambda_{S3}^2.
\eeq
In our analysis, the production cross section of the wino and gluino is crucial.
It is determined by the wino and gluino masses, and almost independent on the sfermion masses and tan $\b$.
We set $\Lambda_{S1}^2=\Lambda_{S2}^2=\Lambda_{S3}^2=2\times10^{11}~\GEV^2$
and $\tan\beta=10$.
Our result does not strongly depend on these values.
We investigate the parameter space $(\Lambda_{G2}, \Lambda_{G3})$.
To calculate the low energy MSSM masses, we have used the program SOFTSUSY 2.0.18 \cite{Allanach:2001kg}.

\begin{figure}[htbp]
\begin{center}
\epsfig{file=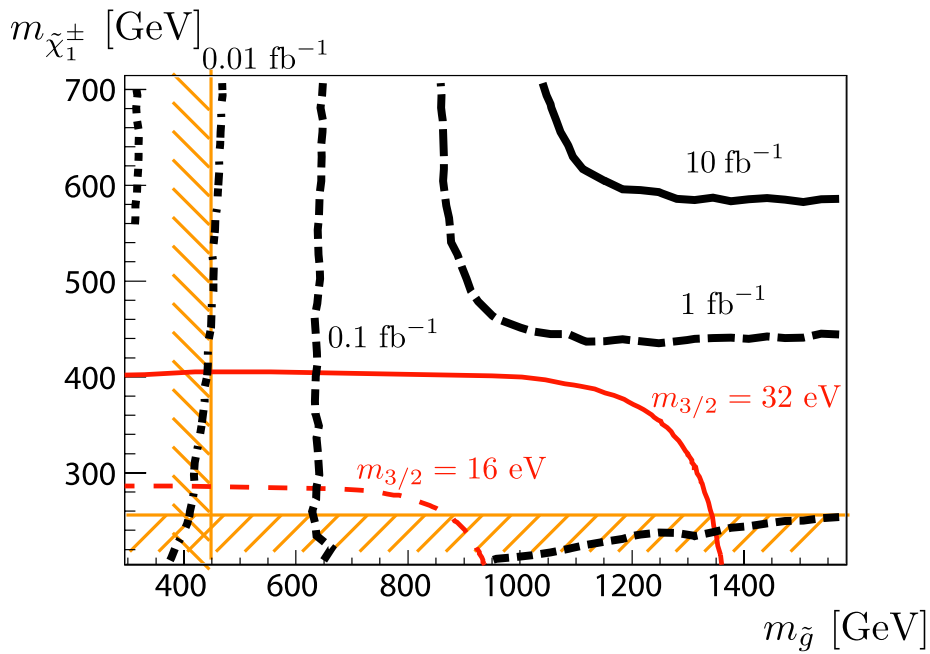 ,scale=0.6,clip}
\end{center}
\caption{Discovery region for $\sqrt{s}=7$ TeV.
The red line represents the allowed region in the present model.
The yellow shaded region is Tevatron exclusion region.
}
\label{fig:7TeV}
\end{figure}

We define the signal significance as
\beq
Sig = \frac{\rm \#~of~signals}{ \sqrt{ \max({\rm \#~of~BG},1) } }.
\eeq
In Fig.~\ref{fig:7TeV}, we show the discovery region where the significance $Sig$ reaches 5.
The integrated luminosity of the LHC will reach $1~{\rm fb}^{-1}$ by the end of 2011.
We can see that the present model with both $m_{3/2}=16$ and 32 eV can be excluded/discovered in the early stage of the LHC.

\section{Conclusion and Discussion}
In this Letter, we investigated the parameter space of the low energy GMSB with a stable vacuum and
have found that this model is completely excluded/discovered in the early stage of the LHC run.

Although we focus on a specific model, the analysis is quite general.
The assumption is the sfermions are heavy enough to be irrelevant at the LHC.
This is a common feature of a large class of non-minimal GMSB models \cite{Shirai:2010rr}.
If the sfermion masses are comparable to the gaugino mass like a minimal GMSB,
the sfermion production can contribute the total SUSY production.
Therefore, it is expected that the total SUSY production cross section is enhanced.
In the case of the minimal GMSB, we have checked that the total SUSY production cross section is enhanced.

Finally, let us comment on the measurement of the SUSY particles.
In low luminosity, it is difficult to measure the SUSY mass precisely.
The cross section gives good estimation of the gluino/wino mass.
As for the gravitino mass, a relation between momentum of photons and missing energy gives
an implication of the gravitino mass \cite{Shirai:2009kn}.

\section*{Acknowledgements}
We would like to thank   T.~T.~Yanagida and K.~Yonekura for useful discussions and careful reading of the manuscript.
This work is supported in part by JSPS
Research Fellowships for Young Scientists and by
World Premier International Research Center Initiative, MEXT, Japan.

\end{document}